\documentclass[pra,aps,twocolumn,superscriptaddress,floatfix]{revtex4-1}
\usepackage{latexsym, amssymb, amsfonts,amsmath, amstext,graphicx, color, bm, times, hyperref, relsize, mathptm, soul,amsthm}
\usepackage[utf8]{inputenc}
\usepackage[english]{babel}
\newcommand{\tr}{\text{tr}}
\def\tr{\mbox{tr}}

{\catcode`\|=\active
  \gdef\Braket#1{\begingroup
\mathcode`\|32768\let|\BraVert\left<{#1}\right>\endgroup}}
\def\BraVert{\egroup\,\mid\,\bgroup}

\definecolor{kmblue}{rgb}{0.19, 0.25, 0.91}
\definecolor{kmred}{rgb}{0.79, 0.29, 0.0}
\definecolor{kmgreen}{rgb}{0, 0.42, 0.24}

\newcommand{\add}[1]{\color{black}#1}

\begin{document}

\title{Unattainable \& attainable bounds for quantum sensors}

\author{Masahito Hayashi}
\affiliation{Graduate School of Mathematics, Nagoya University,
Furocho, Chikusaku, Nagoya, 464-8602, Japan.}
\affiliation{Centre for Quantum Technologies, National University of Singapore, 3 Science Drive 2, Singapore 117543.}
\author{Sai Vinjanampathy}
\email{sai@quantumlah.org}
\affiliation{Centre for Quantum Technologies, National University of Singapore, 3 Science Drive 2, Singapore 117543.}
\author{L. C. Kwek}
\affiliation{Centre for Quantum Technologies, National University of Singapore, 3 Science Drive 2, Singapore 117543.}
\affiliation{Institute of Advanced Studies, Nanyang Technological University, 60 Nanyang View, Singapore 639673.}
\affiliation{National Institute of Education, Nanyang Technological University, 1 Nanyang Walk, Singapore 637616.}

\date{\today}

\begin{abstract}
In quantum metrology, it is widely believed that the quantum Cram\'{e}r-Rao bound is attainable bound while it is not true.
In order to clarify this point, we explain why the quantum Cram\'{e}r-Rao bound cannot be attained geometrically.
In this manuscript,
we investigate noiseless channel estimation under energy constraint for states,
using a physically reasonable error function, and present 
the optimal state and the attainable bound. We propose the experimental generation of the optimal states for enhanced metrology using squeezing transformations. This makes the estimation of unitary channels physically implementable, 
while existing unitary estimation protocols do not work.
\end{abstract}

\maketitle
\section{Introduction}
The theory of estimation is at the heart of \add{modern quantum sensing} \cite{gendra2013quantum,chiribella2012optimal,de2015two,giovannetti2006quantum}. Quantum estimation, \add{which employs quantum states to estimate unknown parameters}, can be divided into state estimation and channel estimation problems. In state estimation problem, the state is parametrized by the unknown parameter, which is estimated by subjecting \add{it} to \add{quantum} measurements. This is contrasted against channel estimation, where \add{a quantum} channel is parametrized by a set of \add{unknown} parameters. Such an estimation problem is approached by choosing a set of optimal input states, and subjecting the states that are output by the channel to measurements. 

 To \add{compare two} designs of quantum sensors, we must \add{employ} a figure of merit that estimates how well a given strategy of parameter estimation is doing. A commonly used figure of merit is the uncertainty of the estimated parameter, denoted by $\Delta\theta$. If we imagine an optical interferometric setup, the so-called \textit{standard quantum limit} refers to  strategies wherein the uncertainty scales inversely with the square-root of the average number of photons in the input state, namely $\Delta\theta\propto\langle N\rangle^{-\frac{1}{2}}$. The chosen figure of merit is often studied under some constraints on the physical system. We note that there are several different constraints of the problem of parameter estimation that have been studied in the literature. The maximum photon number and the average photon number are examples of quantities which have been constrained by metrology schemes \cite{dorner2009optimal,huver2008entangled,hayashi2012fourier,hayashi2011constraint,hayashi2009,PhysRevLett.62.2377}. 
Likewise non-linear Hamiltonians have been investigated and in the context of quantum enhanced metrology \cite{PhysRevA.76.035801,boixo2008quantum,boixo2008quantumPRA}. 

 
\add{Some authors have estimated the uncertainty of estimation by using the noise-to-signal ratio. However, it was noted that this definition is unsatisfactory for multimodal probability distributions \cite{dowling2008quantum}.} 
Another common choice of uncertainty that has been studied is the Cram{\'e}r-Rao bound (CRB)
which specifies the error bound. 
However, it does not work properly in the unitary estimation 
because any estimation strategy cannot saturate 
the CRB while it can be saturated by two-step strategies in the state estimation and the noisy channel estimation \cite{hayashi2011comparison}.
 
In this manuscript, 
we clarify the reason why a two-step strategy can saturate the CRB 
in the state estimation and the noisy channel estimation
and it fails to saturate it in unitary (noiseless) estimation problems.
This discussion shows why the N00N state does not work for unitary estimation in practice
nevertheless its importance is widely believed.
To obtain the truly attainable bound under the energy constraint, 
we convert the minimization problem of the average errors 
to another simple problem in the asymptotic regime.
Using squeezing operation,
as an implementable solution in quantum optical system,
we propose a concrete protocol for parameter estimation
with quadratic enhancement with respect to the constrained energy $E$.
Furthermore, we also propose an experimental scheme to generate the optimal states used in the protocol.

\section{CRB is attainable for state estimation ---}.  CRB relates the uncertainty in the estimate of an unbiased estimator to the so-called Fisher information \cite{helstrom,holevo,durkin2007local}. The quantum CRB reads
\begin{align}
\Delta\theta\geq\frac{1}{\sqrt{\nu}\sqrt{\mathcal{F}_Q}},
\end{align}
where $\mathcal{F}_Q$ is the quantum Fisher information. \add{We emphasize that CRB, and its quantum generalizations, apply only asymptotically, in the limit of infinitely many repetitions $\nu$ of independent measurements.} Whether a calculated asymptotic minimum can be attained via implementable optimal measurements has been investigated by many authors. 
QCRB bound has been often criticized as being an unphysical bound, 
since it is only valid for local estimation. 
That is, the optimal measurement to attain the above bound depends on the true parameter to be estimated.
To overcome this problem, the two-step strategy \cite{HayashiMatsumoto,gill2000state,hayashi2011comparison} is employed. 

The mean-square error associated with state estimation is given by 
\begin{align}
{\rm{MSE}}_{\theta}(M^{(N)}):=\displaystyle\int  (\theta-\theta_{\rm est})^2 \tr(\rho^{\otimes N}_{\theta}M^{(N)}(d\theta_{\rm est})).
\end{align}
The infimum of ${\rm{MSE}}_{\theta}(M^{(N)})$ over all measurements $M^{(N)}$ is a measure of the ultimate precision achievable by the setup described in figure (1.b). 
Since the estimator is assumed to be unbiased, we have to impose the additional condition
\begin{align}
\displaystyle\int \theta_{\rm est} \tr(\rho^{\otimes N}_{\theta}M^{(N)}
(d\theta_{\rm est}))
=\theta,~\forall~\theta\in\Theta.
\end{align}
\add{Here $\Theta$ is the set of allowed values of $\theta$. }This condition is too restrictive, since there are often no measurements that can satisfy this condition of being globally unbiased. Hence, we can modify the condition to be locally unbiased, by demanding that the above equation 
be true with the first order Taylor expansion 
at a given value $\theta=\theta_0$ \cite{holevo}. 

The optimal locally unbiased estimator can be used in a two-step strategy to estimate the unknown parameter such that the MSE saturates the CRB as follows \cite{HayashiMatsumoto,gill2000state}.
In the two-step strategy,
we first get the estimate $\theta_1$ for the unknown global parameter 
by employing the first $l$ copies of the state.
In the second step, we apply the optimal locally unbiased estimator at $\theta_1$
for remaining $N-l$ copies to refine the estimate. 
When $\theta_1$ falls in a ball of radius $\delta$ about the true parameter,
the MSE is sufficiently close to the CRB.
Since the radius $\delta$ does not depend on the numbers $l$ and $N$,
the first estimate $\theta_1$
falls in the ball of radius $\delta$ with almost probability 1 as $l$ is sufficiently large.
Hence, for state estimation, this strategy saturates the CRB.
That is, the CRB is considered to express the ultimate bound for precision of state estimation \cite{HayashiMatsumoto,gill2000state}. This is not true for unitary estimation, as discussed below.

\section{CRB is unattainable for unitary estimation}
In the unitary estimation, Fig.~(1.b), 
to estimate the parameter $\theta$ parameterizing the unknown unitary $U_\theta$
we can choose the input state $|\psi\rangle$ as well as
the measurement $M^{(N)}$.
Hence, we denote the MSE by ${\rm{MSE}}_{\theta}(M^{(N)},|\psi\rangle)$.
When we choose the input state $|\psi\rangle$,
we have state family 
\add{$\{U^{\otimes N}_\theta |\psi\rangle \langle \psi|U_{\theta}^{\dagger\otimes N}\}_\theta $,}
whose the quantum Fisher information at $\theta_0$ 
is denoted by ${\cal F}^{(N)}_{\theta_0}[|\psi\rangle]$.
It is known that N00N states realize the maximum quantum Fisher information
$\max_{|\psi\rangle}{\cal F}^{(N)}_{\theta_0}[|\psi\rangle]$, 
which \add{scales as} $O(N^2)$ \cite{giovannetti2004quantum}.
Hence, it is believed that 
N00N states attain the inverse of this maximum \cite{giovannetti2006quantum}. Even in the case of N00N states, we note that the original scheme for enhanced lithography \cite{boto2000quantum} was criticized as being impractical \cite{tsang2007relationship}. 
In Appendix \ref{a3}, we formally show that the CRB cannot produce any bound for unitary estimation.

However, as shown in \cite{hayashi2011comparison}, the CRB cannot be saturated by any operation, i.e., any pair of input state and measurement.
Since the two-step strategy outlined in Fig~(1.c) seems to saturate it even in unitary estimation,
we demonstrate how the two-step strategy 
cannot achieve the CRB
in the unitary estimation.
Similar to state estimation, the optimal input state and 
the optimal measurement
depend on the true parameter $\theta$, and are denoted by $|\psi_\theta\rangle$
and $M^{(m)}_{\theta}$.
In the first step, we obtain the tentative estimate $\theta_1$ by employing $l$ copies.
Then, the second step is $(M^{(m)}_{\theta_1},\psi_{\theta_1})$,
whose error is evaluated as follows.
For every $\varepsilon>0$ and $m:=N-l$, 
we choose the error bar 
$\delta:=\delta(m,\varepsilon)= C_\epsilon \frac{1}{m}$ 
with a suitable choice of the constant $C_\epsilon$.
Then,
$\vert\theta_1-\theta_0\vert\leq\delta$ implies that 
\begin{align}
m^2\displaystyle MSE^{(m)}_{\theta_0}[M^{(m)}_{\theta_1},\psi_{\theta_1}]
\leq 
m^2 
\left(\max_{|\psi\rangle}
{\cal F}^{(m)}_{\theta_0}[|\psi\rangle]\right)^{-1}
+\varepsilon.
\end{align}
The meaning of the above equation is as follows: if $\theta_1=\theta_0$, 
then the LHS of the equation above achieves the CRB.
The equation is normalized by $m^2$ 
because $MSE\propto m^{-2}$. 
Unfortunately, this condition is satisfied only with 
the probability $\rm{Pr}\{\vert\theta_1-\theta_0\vert < \delta\}$. 
Although $\delta$ behaves as $O(\frac{1}{m})$, 
the error 
$\vert\theta_1-\theta_0\vert$ in the first step 
behaves as $O(\sqrt{\frac{1}{l}})$.
In the limit $N\rightarrow\infty$, 
this probability becomes very small. 
Hence the CRB given above cannot be attained
by the two-step method.
That is, in the unitary estimation, 
the CRB cannot be used as a figure of merit.
We note that when the target to be estimated is noisy channel, 
the above two-step strategy works well \cite{hayashi2011comparison}.

\begin{figure}[t]
\begin{center}
\includegraphics[width=0.35\textwidth]{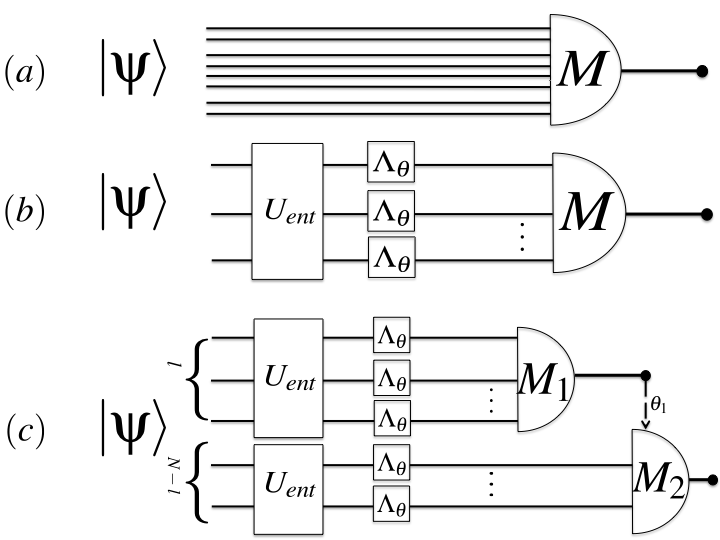}
\caption{\label{Strategy} Three different metrology schemes are presented, namely (a) state estimation problem, where a parametrized state is to be determined, 
(b) single step unitary estimation, 
where an initial state is entangled and then subjected to a unknown unitary, followed by measurements, and 
(c) a two-step strategy where a subset of the the initial states are used to estimate the neighbourhood of the true parameter, 
and the rest of the states are used to perform an adaptive measurement to estimate the unknown parameter $\theta$.}
\end{center}
\end{figure}

To overcome this problem, we need to seek another figure of merit.
Although in the above we have discussed the case when the unknown unitary is given as $n$ copies of 
a given unknown unitary,
we will consider a more physical situation to discuss the parameter estimation problem.
It is natural that 
the unknown parameter is the phase parameter acquired by the Hamiltonian.
In optical system,
the Hamiltonian is given by the number operator $\hat{n}$, 
and the initial state is considered to be $\vert \psi\rangle=\sum_{n=0}^{\infty}\psi_n\vert n\rangle$, whose coefficients we will determine, for optimal performance. 
In order to avoid the ambiguity caused by the periodicity,
we employ the error function $2\sin^{2}(\theta_{\rm est}-\theta_0)$. 
This error function is approximately quadratic for small values and zero when $\theta_{\rm est}$ is close to $\theta_0$. 
The average error is then defined as
\begin{align}
\mathcal{D}(M,\vert\psi\rangle):=2\displaystyle\int_{0}^{2\pi}
\sin^2(\theta_{\rm est}-\theta)\langle\psi\vert e^{i\hat{n}\theta} M(d\theta_{\rm est})e^{-i\hat{n}\theta}\vert\psi\rangle.
\end{align}
For a physically reasonable constraint for input states,
we constrain the energy of the state, namely $\langle\psi\vert \hat{n}\vert\psi\rangle\leq \mathrm{E}$
because it bounds the error function in terms of the energy resources employed in the metrology scheme. 
Energy constraint \cite{giovannetti2004quantum,hall2012heisenberg,giovannetti2006quantum,rivas2012sub,tsang2012ziv} is one of the few important resource constraints  \cite{hayashi2012fourier,hayashi2011constraint,hayashi2009} considered in quantum metrology. With this constraint, the minimum error, which is a good figure of merit for metrological tasks, can be written as
\begin{align}\label{minE}
\tau(\mathrm{E}):=\displaystyle\min_{(M,\vert\psi\rangle)}
\{\mathcal{D}(M,\vert\psi\rangle)\vert\langle\psi\vert \hat{n}\vert\psi\rangle\leq \mathrm{E}\}.
\end{align}
In this scenario, the resource that is varied is the fixed value of $E$ instead of the number of copies.

\section{Attainable bound} 
Using the figure of merit derived above, we now derive an attainable limit in the limit of large energy $E$.
Since the error function has a group covariant form, we can restrict our measurement, without loss of generality, to a covariant measurement \cite{holevo2011probabilistic,d2001using}, namely
\begin{align}
\displaystyle M_0(d\theta_{\rm est}):=e^{-i\hat{n}\theta_{\rm est}}\left[\sum_{n,n'}\vert n\rangle\langle n'\vert \right]e^{i\hat{n}\theta_{\rm est}}\frac{d\theta_{\rm est}}{2\pi}.
\end{align}
 Finally, we notice that $\mathcal{D}(M_0,\vert\psi\rangle)$ can be split into two non-interacting parts $\mathcal{D}(M_0,\vert\psi_e\rangle)$ and $\mathcal{D}(M_0,\vert\psi_o\rangle)$, corresponding to even and odd parity sectors of the state $\vert\psi\rangle$. Here $\vert\psi_e\rangle\propto\sum_{n=0}^{\infty}\psi_{2n}\vert 2n\rangle$ and likewise $\vert\psi_o\rangle\propto\sum_{n=0}^{\infty}\psi_{2n+1}\vert 2n+1\rangle$.

We can hence deal with the parity sectors separately. Such good parity states and operators have already been shown to be advantageous in quantum metrology schemes \cite{plick2010parity,anisimov2010quantum,gerry2010parity,chiruvelli2011parity}. To address the optimization in the asymptotic regime, 
consider a square integrable function $f$ on $\mathbb{R}_+$ with the existence of the limit $\lim_{x\rightarrow+0}f(x)$. 
Such a function will serve as continuous generalizations of the discrete coefficients $\psi_{n}$. 
We then define the input state $\vert\psi_{f,R,e}\rangle=\sum_{n=0}^{\infty}\psi_{f,R,e\vert n}\vert 2n\rangle$. 
Here $\psi_{f,R,e\vert n}:=f(n/R)/\sqrt{R}$ is the function, normalized so that the total probability is still unity. 
Since we have $\langle\psi_{f,R,e\vert n}\vert \hat{n}\vert\psi_{f,R,e\vert n}\rangle=2R\langle f\vert Q\vert f\rangle$ by using the position operator  
$Q:f\rightarrow xf(x)$,
we discuss the asymptotic regime $E \to \infty$ by increasing $R$.
Then, the error function $\mathcal{D}(M_0,\psi_{f,R,e\vert n})$ 
can be simplified to 
\begin{align}
\mathcal{D}(M_0,\psi_{f,R,e\vert n})\approx\frac{\vert f(0)\vert^2}{2R}+\frac{1}{2}\langle f\vert P^2\vert f\rangle
\end{align}
by using the momentum operator $\hat{P}(f(x)):f\rightarrow idf(x)/dx$. We note that the first term can be ignored in the limit of large $R$ if $f(0)=0$, a condition that will be employed below. This means that 
\begin{align}
\langle\psi_{f,R,e\vert n}\vert \hat{n}\vert\psi_{f,R,e\vert n}\rangle^2\mathcal{D}(M_0,\psi_{f,R,e\vert n})\approx2\langle f\vert Q\vert f\rangle^2\langle f\vert P^2\vert f\rangle.
\end{align}

A similar equation can be shown to hold true for the odd states. While \cite{A,B} showed $\tau(\mathrm{E})=\mathcal{O}(1/\mathrm{E}^2)$, taking these two cases into account, we can asymptotically characterize the minimum error $\tau(\mathrm{E})$ as 
\begin{align}\label{basic_form}
\mathrm{E}^2\tau(\mathrm{E})\approx \min_{f:f(0)=0}2\langle f\vert Q\vert f\rangle^2\langle f\vert P^2\vert f\rangle\geq\frac{1}{8},
\end{align}
where the lower bound is shown to hold in Appendix \ref{a2}. 
In the optimal case, we can hence realize the quadratic enhancement where the error scales inversely with the square of the energy. 
Hence, the minimum value on the right hand side gives the truly attainable bound instead of the CRB.
Since this optimization is difficult in general, we handle the optimization program by restricting ourselves to the class of square integrable functions $\{\psi_a\}_{a>0}$ with $\psi_{a}(x):=x^ae^{-x/2}/\sqrt{\Gamma(1+2a)}$, where $\Gamma(x)$ is the gamma function \cite{abramowitz1964handbook} and seek an analytical solution. 
Such a form for $\psi_{a}(x)$ is motivated from the facts that (a) the functional value at $x=0$ should be zero, and (b) the function should be square integrable and normalized to unity. In this case, the two quantities $\langle\psi_a\vert Q\vert\psi_a\rangle=2a+1$ and $\langle\psi_a\vert P^2\vert\psi_a\rangle=[4(2a-1)]^{-1}$ can be evaluated. The second term is made finite by demanding that $a>1/2$. Now, we obtain an attainable error coefficient, which is the product prescribed in Eq.(\ref{basic_form}) and is calculated to be $c(a)=(2a+1)^2/[2(2a-1)]$.
The minimum value, for the constraint $a>1/2$, is realized at $a=3/2$. 
 
\section{Input State Generation via Squeezing}

Below, we consider 
the suboptimization accompanying squeezing transformations and its physical generation, which 
are implementable as the metrological scheme.
This is because squeezing transformations are implemented routinely in quantum optics experiments.
By using the annihilation and creation operators $a$ and 
$a^{\dagger}$,
the squeezing operator is given as 
$S(\xi):=exp([\xi^{*} a^2-\xi a^{\dagger^2}])$, where
$\xi$ is related to the squeezing parameter $0\leq r<\infty$ by the relation $\xi=r\exp(i\varphi)$, with $0\leq\varphi<2\pi$. 
We optimize the input state among superposition of squeezed number states 
because several authors have considered superpositions of squeezed states with other states, such as thermal states \cite{vourdas1987photon} and squeezing applied to superpositions of coherent states. 
Such studies have focused on the non-classical properties of the resulting quantum states, and their applications to quantum metrology \cite{hofmann2007high}. 
The motivation for considering superpositions of squeezed states follows from the parity separation of the optimal states
because squeezing transformation commutes with the projection operators on both the even and odd sectors \cite{brif1996parity}. 

As shown in Appendix \ref{a5}, 
when we increase the constraint $E$ by increasing the parameter $r$,
the state $|\psi_{a}\rangle$ can be realized 
by superpositions of squeezed photon number states
if and only if $a=l\pm1/4$.
The minimum of \eqref{basic_form}, 
when we restrict ourselves to such a subset, 
is achieved at $a=2-1/4=7/4$. 
We note that such a state is given by $\vert\psi_{7/4}\rangle:=S(r)(c_{0}\vert0\rangle+c_{2}\vert2\rangle+c_{4}\vert4\rangle)$, where $c_0=\sqrt{3/35}$, $c_2=-\sqrt{24/35}$ and $c_4=\sqrt{8/35}$. 

Next, we show how to generate the suboptimal state $\vert\psi_{7/4}\rangle$ we have derived here
by using squeezing transformations \cite{gerry2005introductory}. 
Since both single-mode and two-mode squeezing transformations are routinely implemented in laboratories,
 these transformations can produce the suboptimal state $\vert\psi_{7/4}\rangle$.  
Such states can be produced by first producing a superposition of states, using two-mode squeezed vacuum states as shown below, followed by injection of these states into a degenerate parametric down-converter that implements the squeezing transformation on arbitrary input states.

\begin{figure}[t]
\begin{center}
\includegraphics[width=0.35\textwidth]{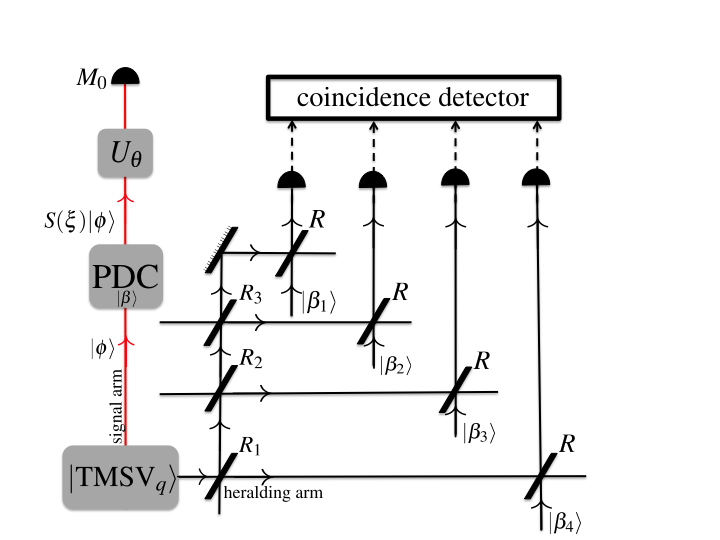}
\caption{\label{generating_State} To generate the optimal state, one of the modes of a two-mode squeezed vacuum state (TMSV) is split four-ways by three beamsplitters with reflectivities $R_1=\sqrt{3}/2$, $R_2=1/\sqrt{3}$ and $R_3=1/\sqrt{2}$. The corresponding outputs are displaced with coherent light of amplitudes $\beta_i$ incident on $R\approx1$. A four-way single photon coincidence detection is performed at the output. The other mode of the TMSV, heralded by this four-way coincidence detection, is input into a degenerate spontaneous parametric down converter, pumped by a non-depleting coherent state $\vert\beta\rangle$. The output, for an appropriately chosen parameters such that $\xi=2\chi^{(2)}\beta t$, is the desired input state. It then acquires the parameter and is measured by $M_0$.}
\end{center}
\end{figure}

The approximate asymptotically suboptimal state is produced by standard laboratory techniques \cite{yukawa2013generating}. Consider a two-mode squeezed vacuum state $\vert\text{TMSV}_q\rangle=\sqrt{1-q^2}\sum_nq^n\vert n\rangle_s\otimes\vert n\rangle_h$, where $s$ denotes signal mode, $h$ denotes heralding mode and $q=\tanh(r_2)$ is related to the two-mode squeezing $r_2$ (different from the single mode squeezing $r$). By splitting the heralding mode into four beams, displacing each beam by $\beta_i$ and measuring for a single photon coincidence in all four beams, we get the output state to be
\begin{align}
\vert\phi\rangle\propto\langle0\vert_h(\frac{a}{2}+\beta_1)(\frac{a}{2}+\beta_2)(\frac{a}{2}+\beta_3)(\frac{a}{2}+\beta_4)\vert\text{TMSV}_q\rangle.
\end{align}
This is depicted in Fig.~(\ref{generating_State}). This state can be written as $\vert\phi\rangle=N^{-1}\sum_{k=0}^{4}\phi_k\vert k\rangle$, where $\phi_0:=\beta_1\beta_2\beta_3\beta_4$, $\phi_1:=\frac{q}{2}(\beta_1\beta_2\beta_3+\beta_1\beta_2\beta_4+\beta_1\beta_3\beta_4+\beta_2\beta_3\beta_4)$, $\phi_2:=\frac{q^2\sqrt{2}}{4}(\beta_1\beta_2+\beta_1\beta_3+\beta_1\beta_4+\beta_2\beta_3+\beta_2\beta_4+\beta_3\beta_4)$, $\phi_3:=\frac{q^3\sqrt{3!}}{8}(\beta_1+\beta_2+\beta_3+\beta_4)$ and $\phi_4:=\frac{q^4\sqrt{4!}}{2^4}$. Now, let us make the choice $\beta_1\beta_2=\beta_3\beta_4$, $(\beta_1+\beta_2)=-(\beta_3+\beta_4)$ since this eliminates the odd parity sectors. The suboptimal state is generated for a modest squeezing of $r_2=2$, with parameters $\beta_1\beta_2\approx0.890702$, $\beta_1+\beta_2\approx2.9344$ and $N\approx2.73989$ and fidelity to the suboptimal state $99.94\%$. This amounts to setting $\beta_1=-\beta_3=0.343824$ and $\beta_2=-\beta_4=2.59058$ at an experim
 entally 
 feasible  choice of two-mode squeezing, namely $r_2=2$.

Once this desired state $\vert\phi\rangle$ is produced, it is input into a degenerate parametric down-converter, which is pumped by a strong laser field $\vert\beta\rangle$. The pump is assumed to be in the undepleted regime. The resulting interaction Hamiltonian $H_I=i\hbar(\eta a^{2}-\eta^* a^{\dagger^2})$, with $\eta=\chi^{(2)}\beta$. Here, $\chi^{(2)}$ stands for the non-linear susceptibility. This squeezes $\vert\phi\rangle$, to produce the suboptimal state $\vert\psi_{{7}/{4}}\rangle$ for the choice $r=\eta t$. Finally, we note that the covariant measurement required for optimal 
metrology can be implemented by coupling to a continuous variable mode.
The detail description and the analysis are included in Appendix \ref{a7} for completeness. 
We also note that a discrete version of covariant measurements is implementable by coupling the light mode to several qubits and performing measurements on the resulting qubit states \cite{cleve1998quantum}.

In this manuscript, presented the optimization of input states for noiseless channel estimation with the cost function described by Eq.~(\ref{minE}). Such a state optimization was simplified by using asymptotic analysis and a suitable optimal state, namely $\vert\psi_{{7}/{4}}\rangle$ was determined. Such a state shows quadratic enhancement in the mean squared error of the estimated parameter. Furthermore, an experimental scheme to generate the suboptimal state was presented. Such a scheme is easily implementable in current quantum optics experiments, paving way for implementable quantum metrology of unitary channels. 

We emphasize that the importance of the current work stems from the fact that CRB is a widely accepted bound in quantum metrology. Though the CRB may often be saturated for noisy channel estimation, the case of unitary estimation poses a unique challenge \cite{hayashi2011comparison}. In this important case of unitary estimation, the CRB is widely accepted as the bound though it does not provide an actual bound for the uncertainty in the estimated parameter. Furthermore, the two-step strategy, which has also been widely accepted as a valid estimation technique, does not work in the case of unitary estimation. In the current manuscript, we clarify the bound for such metrological schemes and suggest an implementable scheme to saturate such a bound.

\begin{acknowledgments} Centre for Quantum Technologies is a Research Centre of Excellence funded by the Ministry of Education and the National Research Foundation of Singapore.  \add{This research is supported by the National Research Foundation Singapore under its Competitive Research Programme (CRP Award No. NRF-CRP14-2014-02).} MH is partially supported by a MEXT Grant-in-Aid for Scientific Research (A) No. 23246071 and the National Institute of Information and Communication Technology (NICT), Japan.
\end{acknowledgments}

\appendix

\section{Cost Function and Parity Sectors}\label{a1}
To show Eqs. (8) and (10) in the main text, 
we prepare several mathematical notations.
The average error in Eq.~(5) is given by
\begin{align}
\mathcal{D}(M,\vert\psi\rangle):={2}\displaystyle\int_{0}^{2\pi}
\sin^2(\theta_{\rm est}-\theta)\langle\psi\vert e^{i\hat{n}\theta} M(d\theta_{\rm est})e^{-i\hat{n}\theta}\vert\psi\rangle.
\end{align}
For a given amount of maximum energy $E$, we consider the following cost function
\begin{align}\label{minE}
\tau(\mathrm{E}):=\displaystyle\min_{(M,\vert\psi\rangle)}
\{\mathcal{D}(M,\vert\psi\rangle)\vert\langle\psi\vert \hat{n}\vert\psi\rangle\leq \mathrm{E}\}.
\end{align}
Since the error function has the group covariant form, 
we can restrict our measurement into covariant measurement.
So, the measurement $M$ is written as 
\begin{align}
\displaystyle M(d\theta_{\rm est})
:=e^{-i\hat{n}\theta_{\rm est}}
\left(
\sum_{n,n'}e^{ia_n} \vert n\rangle\langle n'\vert e^{-ia_n} 
\right)
e^{i\hat{n}\theta_{\rm est}}\frac{d\theta_{\rm est}}{2\pi}.
\end{align}
Further, without loss of generality, 
our measurement can be restrict to the following POVM
in the following sense
\begin{align}
\displaystyle M_0(d\theta_{\rm est}):=e^{-i\hat{n}\theta_{\rm est}}\left[\sum_{n,n'}\vert n\rangle\langle n'\vert \right]e^{i\hat{n}\theta_{\rm est}}\frac{d\theta_{\rm est}}{2\pi}.
\end{align}
That is, 
the error ${\cal D}(M,\vert\psi\rangle)$
equals 
${\cal D}(M_0,\vert\psi'\rangle)$ with
$\vert\psi'\rangle:=\sum_{m=0}^{\infty}\psi_m e^{-ia_m}\vert m\rangle$,
which can be shown as
\begin{align}
&{\cal D}(M,\vert\psi\rangle) \nonumber\\
=&
\int_{0}^{2\pi} {2}\sin^2(\theta_{\rm{est}}-\theta)
\langle \psi\vert 
e^{i \hat{n} (\theta_{\rm{est}}-\theta)}
\nonumber\\
& \cdot 
\bigg(
\sum_{n,n'} e^{ia_n}\vert n\rangle \langle n'\vert e^{-ia_{n'}}
\bigg) 
e^{-i \hat{n} (\theta_{\rm{est}}-\theta)}
 \vert\psi\rangle 
\frac{d\theta_{\rm{est}}}{2\pi} \nonumber\\
=&
\int_{0}^{2\pi}{2}\sin^2(\theta_{\rm{est}}-\theta)
\bigg(
\sum_{m} \langle m\vert  {\psi^{*}_m e^{ia_m}}
\bigg)
e^{i \hat{n} (\theta_{\rm{est}}-\theta)}
\nonumber\\
&\cdot 
\bigg(
\sum_{n,n'} \vert n\rangle \langle n'\vert 
\bigg)
e^{-i \hat{n} (\theta_{\rm{est}}-\theta)}
\bigg(
\sum_{m'} \psi_{m'} e^{-ia_{m'}}\vert  m' \rangle 
\bigg)
\frac{d\theta_{\rm{est}}}{2\pi} \nonumber\\
=&{\cal D}(M_0,\vert\psi'\rangle).
\end{align}
Since the state $\vert\psi'\rangle$ has the same average energy as 
$\vert\psi\rangle$, 
without loss of generality, we can restrict our measurement to $M_0$. 
That is, it is enough to consider the optimization for the input state, i.e., 
\begin{align}
\tau(E)=\min_{\psi}\{ {\cal D}(M_0,\psi) \vert
\langle \psi \vert \hat{n} \vert\psi\rangle \le E \}.
\end{align}
When the input state is 
$\vert\psi\rangle:= \sum_{n=0}^{\infty} \psi_n\vert n\rangle$,
the average error is calculated to
\begin{align}
&{\cal D}(M_0,\psi) \nonumber \\
=&
\int_{0}^{2\pi} 
(1-\frac{(e^{i2(\theta_{\rm{est}}-\theta)}+e^{-i2(\theta_{\rm{est}}-\theta)})}{2})
\nonumber\\
&\cdot \bigg(\sum_{n}
{\psi^{*}_n}
e^{i n(\theta_{\rm{est}}-\theta)}
\bigg)
\bigg(\sum_{n'}
\psi_{n'}
e^{-i n'(\theta_{\rm{est}}-\theta)}\bigg)
\frac{d\theta_{\rm{est}}}{2\pi} \nonumber\\
=&
1-
\sum_{n=0}^{\infty}
\frac{1}{2}
({\psi^{*}_n} \psi_{n+2}
+ {\psi^{*}_{n+2}} \psi_{n}) \nonumber\\
=& \lambda{\cal D}(M_0,\psi_e)
+(1-\lambda){\cal D}(M_0,\psi_o).
\end{align}
Here
$\vert\psi_e\rangle:= \frac{1}{\sqrt{\lambda}}\sum_{n=0}^{\infty} \psi_{2n}\vert 2n\rangle$,
$\vert\psi_o\rangle:= \frac{1}{\sqrt{1-\lambda}}
\sum_{n=0}^{\infty} \psi_{2n+1}\vert 2n+1\rangle$,
and $\lambda := \sum_{n=0}^{\infty} |\psi_{2n}|^2$. This proves our claim that the average error is not affected by the correlations between the parity sectors, and hence the photon number states with fixed parity can be treated independently.

\section{Optimization in the Asymptotic Regime}\label{a2}
Now that we have discussed parity sectors, let us discuss the optimization of the cost function with respect to the initial state in the asymptotic regime to show Eqs. (8) and (10) in the main text.
Let us parametrize the even state as $\vert\psi_{f,R,e}\rangle:=\sum_{n=0}^{\infty}\psi_{f,R,e\vert n}\vert2n\rangle$ where the subscript $e$ stands for the even sector, the subscripts $f,R,e$ stand for the transformation 
\begin{align}
\psi_{f,R,e\vert n}:= f(n/R)/\sqrt{R}. 
\label{M1}
\end{align}
This is a functional representation of the coefficient such that the corresponding state probabilities add to 1. This is seen from the fact that $f(n/R)/\sqrt{R}$ is square integrable with variable $x=n/R$. Using the standard transformation of the momentum operator $\hat{P}\rightarrow id/dx$, we can write
\begin{align}
&\mathcal{D}(M_0,\psi_{f,R,e}) \nonumber \\
=&1-\displaystyle\sum_{n=0}^{\infty}\frac{\psi^{*}_{f,R,e\vert n}\psi_{f,R,e\vert n+1}+\psi_{f,R,e\vert n}\psi^{*}_{f,R,e\vert n+1}}{2}\nonumber\\
=&\frac{1}{2}\vert\psi_{f,R,e\vert0}\vert^2 \nonumber \\
& +\displaystyle\sum_{n=0}^{\infty}\frac{1}{2}\left(\psi^{*}_{f,R,e\vert n+1}-\psi^{*}_{f,R,e\vert n}\right)\left(\psi_{f,R,e\vert n+1}-\psi_{f,R,e\vert n}\right) \nonumber\\
\cong &\frac{1}{2R}\vert f(0)\vert^2+\frac{1}{2R^2}\displaystyle\int_{-\infty}^{\infty}dx\left\vert\frac{df(x)}{dx}\right\vert^2 \nonumber \\
=& \frac{1}{2R}\vert f(0)\vert^2+
\frac{1}{2R^2}\langle f\vert \hat{P}^2\vert f\rangle\label{last14}.
\end{align}
So, we obtain Eq (8) in the main text.
We will demand that $f(0)=0$, to fix the first term of Eq.~(\ref{last14}). This causes the average error to go to zero quadratically as $1/R^2$ as claimed. The average energy in this case is given by
\begin{align}
\langle\psi_{f,R,e}\vert \hat{n} \vert\psi_{f,R,e}\rangle&=\displaystyle\sum_{n=0}^{\infty}2n\left\vert\frac{1}{\sqrt{R}}f(\frac{n}{R})\right\vert^2\nonumber\\
&=2R\displaystyle\sum_{n=0}^{\infty}\frac{n}{R^2}\left\vert f(\frac{n}{R})\right\vert^2\nonumber\\
&\cong 2R\int_{0}^{\infty}dx x\vert f(x)\vert^2=2R\langle f\vert \hat{Q}\vert f\rangle.
\end{align}
We can hence write
\begin{align}
\langle \psi_{f,R,e}\vert \hat{n} \vert \psi_{f,R,e}\rangle^2\mathcal{D}(M_0,\psi_{f,R,e})
=2\langle f\vert \hat{Q} \vert f\rangle^2\langle f\vert \hat{P}^2\vert f\rangle.
\end{align}
The same can be verified for an odd parity state
$|\psi_{f,R,o}\rangle=\sum_{n=0}^{\infty}\psi_{f,R,o|n}|2n+1\rangle$ 
with 
$\psi_{f,R,o|n}:=f(n/R)/\sqrt{R}$ and $D(M_0,\psi_{f,R,o})=\vert f(0)\vert^2/2R+\langle f\vert \hat{P}^2\vert f\rangle/(2R^2)$ and $\langle \psi_{f,R,o}\vert \hat{n} \vert \psi_{f,R,o}\rangle=1+2R\langle f\vert \hat{Q} \vert f\rangle$. Hence, we can write the assertion used as Eq. (10) in the main text, namely
\begin{align}
E^2\tau(E)\rightarrow\displaystyle
\min_{f:f(0)=0}\big[2\langle f\vert \hat{Q} \vert f\rangle^2\langle f\vert \hat{P}^2\vert f\rangle\big].\label{5-4-2}
\end{align}
To handle the above optimization, we restrict ourselves to a class of functions $\{\psi_a\}_{a>0}$ with $\psi_{a}(x):=x^ae^{-\frac{x}{2}}/\sqrt{\Gamma(2a+1)}$. Using this functional form, we can compute the cost function to be $c(a):=2\langle\psi_a\vert \hat{Q}\vert\psi_a\rangle^2
\langle \psi_a\vert \hat{P}^2\vert \psi_a\rangle
=(2a+1)^2/2(2a-1)$. Restricting $a>1/2$ avoids the singularity, and the minimum cost is given by $c(3/2)=1$. We show below, that the alternative choice of $a=7/4$ produces a near optimal cost $c(7/4)=81/80$ and the corresponding optimal state is given by $\vert \psi_{7/4}\rangle$ which can be produced by squeezing transformation.

Now, we consider the lower bound of \eqref{5-4-2}.
We apply the relation
\begin{align}
\langle f |A^2|f \rangle
\langle f |B^2|f \rangle
\ge
\frac{1}{4}|\langle f |[A,B]|f \rangle|^2
\end{align}
to the case when $A=P$ and $B=Q^{1/2}$.
Thus,
\begin{align}
\langle f |P^2|f \rangle
\langle f |Q|f \rangle
\ge
\frac{1}{4}\langle f |\frac{1}{2}Q^{-1/2}|f \rangle^2
\end{align}
Since the function $x^{-1/2}$ is convex function, we have
$
\langle f |Q^{-1/2}|f \rangle
\ge 
\langle f |Q|f \rangle^{-1/2}$.
Thus,
\begin{align}
\langle f |P^2|f \rangle
\langle f |Q|f \rangle
\ge
\frac{1}{16}(\langle f |Q|f \rangle^{-1/2})^2,
\end{align}
which implies that
\begin{align}
2
\langle f |Q|f \rangle^2
\langle f |P^2|f \rangle
\ge
\frac{1}{8}
\end{align}
Therefore,
\begin{align}
\lim_{E \to \infty}E^2\tau(E)
\ge \frac{1}{8}.\label{5-5-1}
\end{align}

\section{Comparison with Cramer-Rao bound}\label{a3}
Now, we compare our bound 
$\min_{f:f(0)=0}\big[2\langle f\vert \hat{Q} \vert f\rangle^2\langle f\vert \hat{P}^2\vert f\rangle\big]$
with the Cramer-Rao bound (CRB).
When the input state is $|\psi\rangle \langle \psi|$,
we have the family of pure states
$\{ e^{-i \hat{n} \theta} |\psi\rangle \langle \psi| e^{i \hat{n} \theta}
\}$.
The symmetric logarithmic derivative (SLD) Fisher information is 
$J_\psi:=4 \| -i\hat{n}\psi -  \langle \psi|-i\hat{n}\psi\rangle \psi \|^2
=4 (\langle \psi| \hat{n}^2|\psi\rangle
-\langle \psi| \hat{n}|\psi\rangle^2)$.
That is, the CRB is given as $1/J_\psi$.
However, our error is $2 \sin^2 (\theta_{est}-\theta)
\cong 2(\theta_{est}-\theta)^2$ when $\theta_{est}$ is close to $\theta$.
So, to adjust to our error criterion,
we employ the modified Cramer-Rao bound (MCRB)
$2/J_\psi$.
Now, we take account into our energy constraint as
$\langle \psi |\hat{n}|\psi \rangle \le E$.
Thus, we obtain the error bound
$\min_{\psi:\langle \psi |\hat{n}|\psi \rangle \le E}
2/J_\psi$.
This problem is essentially the same as the maximization of the variance under the fixed average on $\{0,1,2,\ldots, n, \ldots\}$.
However, the maximum is infinity as follows.
Given a parameter $t>0$, we choose
the distribution $P$ as
\begin{align}
P(n)=
\left\{
\begin{array}{cl}
1-\frac{1}{t} & \hbox{ when } 0 \\
\frac{1}{t} & \hbox{ when } \lfloor Et \rfloor\\
0 & \hbox{ otherwise.}
\end{array}
\right.
\end{align}
The variance is $\frac{\lfloor Et \rfloor^2}{t}(1-\frac{1}{t})$,
which goes to infinity as $t$ goes to infinity.
That is,
\begin{align}
\inf_{\psi:\langle \psi |\hat{n}|\psi \rangle \le E}
2/J_\psi
=0
\end{align}
Even with our energy constraint, 
we can realize infinite Fisher information.
However, our minimum error is bounded as \eqref{5-5-1}.
So, the MCRB cannot be attained even under our energy constraint. Similar results were derived in \cite{Tsang15}.

\section{Implementation of POVM $M_0$}\label{a4}
Before proceeding to superposition of squeezed states, 
we consider how to physically realize the POVM $M_0$.
For this purpose, in the phase space, we prepare a wave function $|\varphi\rangle$ whose support is included in $[-\frac{1}{2},\frac{1}{2})$.
For a given state $|\psi\rangle:=\sum_{n}\psi_n|n\rangle$, 
we set the initial state $|\psi\rangle\otimes |\varphi\rangle$.
Then, we apply the unitary evolution 
$e^{i\frac{\pi}{2} H_I}e^{i \hat{n} \hat{P}}$,
where 
$H_I:= \sum_{n} (|n\rangle \langle 0|+|0\rangle \langle n|)\otimes \hat{1}_{[n-\frac{1}{2},n+\frac{1}{2})}$
and $\hat{1}_{[n-\frac{1}{2},n+\frac{1}{2})}$ is the projection to the space of functions  with the support $[n-\frac{1}{2},n+\frac{1}{2})$.
So, the resultant state is 
$ |0\rangle \otimes \sum_n \psi_n \varphi_n$,
where $\varphi_n(x):= \varphi(x-n)$.
Then, we measure the momentum operator $\hat{P}$ in the phase space.
We denote the POVM by $M_1$, and
the outcome $p$ is subject to the distribution 
$|\hat{\psi}(p) |^2 d p$, 
where 
$\hat{\psi}(p)$ is given 
as follows.
\begin{align}
&\hat{\psi}(p)
:= 
\int_{-\infty}^{\infty} e^{i x p} 
\sum_n \psi_n \varphi_n(x)
\frac{d x}{\sqrt{2\pi}}  \nonumber\\
=&
\sum_{n=0}^{\infty}
\psi_n \int_{n-1/2}^{n+1/2}e^{ix p} 
\frac{d x}{\sqrt{2\pi}}  \nonumber\\
=&
\sum_{n=0}^{\infty}
\psi_n e^{in p} 
\int_{-1/2}^{1/2}e^{ixp} \varphi(x) \frac{d x}{\sqrt{2\pi}}  \nonumber\\
=&
\Big(\sum_{n=0}^{\infty}
\frac{\psi_n}{\sqrt{2\pi}} e^{in p} 
\Big)
\Big(\int_{-1/2}^{1/2}e^{ixp} \varphi(x) d x
\Big).
\end{align}
When we replace the state $|\psi\rangle$ by 
$|\phi_{\theta}\rangle:= \sum_n\psi_n e^{-in \theta}|n\rangle$,
the outcome distribution is
$|\hat{\psi}_{\theta}(p) |^2 d p$ with
\begin{align}
\hat{\psi}_{\theta}(p)
=&
\Big(
\sum_{n=0}^{\infty} \frac{\psi_n}{\sqrt{2\pi}} e^{in ( p-\theta )}\Big) 
\Big(\int_{-1/2}^{1/2}e^{ixp} \varphi(x) d x
\Big).
\end{align}
At the first glance, the above measurement $M_1$ seems to have no covariant structure
and have a different statistics from the measurement $M_0$.
However, 
these required properties can be recovered by the following post data processing.

Since the outcome $p$ of the measurement $M_1$ runs over $\mathbb{R}$,
it is suitable to take the modular arithmetic for $2\pi$.
Then, we set the final outcome $\theta_{\rm{est}}$ to be
the modular arithmetic of $p$ for $2\pi$, 
and denote the measurement with the outcome $\theta_{\rm{est}}$ by $M_2$.
So, the probability to obtain $\theta_{\rm{est}}$ is 
\begin{align}
&\sum_{k=-\infty}^{\infty}
\big|\hat{\psi}_{\theta}(\theta_{\rm{est}}+ 2k \pi )\big|^2 \nonumber \\
=&
\sum_{k=-\infty}^{\infty}
\bigg|
\bigg(
\sum_{n=0}^{\infty} \frac{\psi_n}{\sqrt{2\pi}} e^{in ( \theta_{\rm{est}}-\theta )}\bigg) 
\Big(\int_{-1/2}^{1/2}e^{ix(\theta_{\rm{est}}+ 2k \pi )} \varphi(x) d x
\Big)
\bigg|^2
\nonumber\\
=&
\bigg|
\sum_{n=0}^{\infty} \frac{\psi_n}{\sqrt{2\pi}} e^{in ( \theta_{\rm{est}}-\theta )}\bigg|^2
\sum_{k=-\infty}^{\infty}
\bigg|
\int_{-1/2}^{1/2}e^{ix(\theta_{\rm{est}}+ 2k \pi )} \varphi(x) d x
\bigg|^2
\nonumber\\
=&
\Big|
\sum_{n=0}^{\infty} \frac{\psi_n}{\sqrt{2\pi}} e^{in ( \theta_{\rm{est}}-\theta )}\Big|^2,
\end{align}
where the final equation is shown as follows.
Since
$
\alpha_k:=
\int_{-1/2}^{1/2}e^{ix(\theta_{\rm{est}}+ 2k \pi )} \varphi(x) d x
=
\int_{-1/2}^{1/2}e^{i2 k x\pi } e^{ix\theta_{\rm{est}}} \varphi(x) d x
$ can be regarded as the Fourier series of $f(x)=e^{ix \theta_{\rm{est}}} \varphi(x) $,
we have $
\sum_{k=-\infty}^{\infty}
\bigg|
\int_{-1/2}^{1/2}e^{ix(\theta_{\rm{est}}+ 2k \pi )} \varphi(x) d x
\bigg|^2
=
\sum_{k=-\infty}^{\infty}|\alpha_k|^2=1$.
Therefore, we find that 
the measurement $M_2$ has the same statistics as the measurement $M_0$.
That is, we can implement the optimal POVM $M_0$ in the above method.

Now, we consider a more realistic case, i.e., the case when 
the support $\varphi$ is not necessarily included in $[-\frac{1}{2},\frac{1}{2})$.
In this case, we can expect that the probability
$\langle \varphi |\hat{1}_{[-\frac{1}{2},\frac{1}{2})}| \varphi\rangle $ 
is enough close to 1, say $1-\epsilon$.
In this case, the fidelity between the real state $|\varphi\rangle$
and the ideal state $\frac{1}{\sqrt{1-\epsilon}}\hat{1}_{[-\frac{1}{2},\frac{1}{2})})| \varphi\rangle $ is $\sqrt{1-\epsilon}$.
That is, the trace norm distance between two state is less than $\sqrt{\epsilon}$ \cite[(6.106)]{hayashi2014introduction}.
So, the variational distance between 
the the distributions of the real outcome and the ideal outcome is also less than 
$\sqrt{\epsilon}$.
Therefore, in practical, it is enough to realize the initial state in the phase space approximately.

\section{Superposition of Squeezed States}\label{a5}
Now, as mentioned in the main text, 
we show that $|\psi_a\rangle$ can be asymptotically realized
by superpositions of squeezed photon number states if and
only if $a = l \pm \frac{1}{4}$. 
To see how squeezed states are related to our suboptimal solution, let us consider squeezed number states, namely $\vert S_{n,r}\rangle:=S(r)\vert n\rangle:=\exp(-r[a^2-a^{\dagger2}]/2)\vert n\rangle$. 

Using these squeezed number states,
we define the superposition of squeezed number states up to 
the number $2m$ or $2m+1$ as
\begin{align}\label{alpha2m}
\displaystyle\vert\alpha_{2m,r}\rangle
:=& C_{2m,r}\sum_{l=0}^{\infty}(-1)^{m-l}{m\choose{l}}\frac{2^ll! \coth^l(r)}{\sqrt{(2l)!}}\vert S_{2l,r}\rangle,\\
\displaystyle\vert\alpha_{2m+1,r}\rangle
:= & C_{2m+1,r}\sum_{l=0}^{\infty}(-1)^{m-l}{m\choose{l}}
\frac{2^ll!\coth^l(r)}{\sqrt{(2l+1)!}}\vert S_{2l+1,r}\rangle,
\end{align}
where
\begin{align}
C_{2l,r}:=&\frac{(2l)!}{2^{2l}l!}\sqrt{\frac{\Gamma(1/2)}{\Gamma(l+1/2)}}\left(\frac{l!}{\sinh^{2l}r}+\frac{\Gamma(2l+1/2)}{\Gamma(l+1/2)}\right)^{-\frac{1}{2}},\\
C_{2l+1,r}:=&\frac{(2l+1)!}{2^{2l}l!}\sqrt{\frac{\Gamma(1/2)}{\Gamma(l+1/2)(1+\sinh^2(r))}}\nonumber \\
& \cdot \left(l!(2l+1)\sinh^{-2l}r+\frac{\Gamma(2l+1/2)}{\Gamma(l+1/2)}(4l+1)\right)^{-\frac{1}{2}}.
\end{align}
We can choose coefficients $\alpha_{2l,r\vert n}$ and $\alpha_{2l+1,r\vert n}$ as
\begin{align}
\displaystyle\vert \alpha_{2l,r}\rangle &=\sum_{n=l}^{\infty}(-1)^{n+l}\alpha_{2l,r\vert n}\vert 2n\rangle.\\
\displaystyle\vert \alpha_{2l+1,r}\rangle &=\sum_{n=l}^{\infty}(-1)^{n+l}\alpha_{2l+1,r\vert n}\vert 2n+1\rangle.
\end{align}
Such states can be converted to even and odd parity states, and is presented in the next section for completeness. Choosing $x:=n/\sinh^2(r)$, we get 
\begin{align}
\alpha_{2l,r\vert n}\cong\frac{1}{\sinh(r)}\psi_{l-\frac{1}{4}}(x).\label{eq1}\\
\alpha_{2l+1,r\vert n}\cong\frac{1}{\sinh(r)}\psi_{l+\frac{1}{4}}(x).\label{eq2}
\end{align}
These above equations hold true in the limit of large $r$. Hence the asymptotic performance of the proposed metrological scheme is achievable with the states $\psi_{l\pm\frac{1}{4}}$ using superpositions of squeezed number states. In perticular, $\vert\alpha_{4,r}\rangle$ can realize the optimal performance among the various choices of $\psi_{l\pm\frac{1}{4}}$, and is given by
\begin{align}
&\vert\alpha_{4,r}\rangle \nonumber \\
=&\sqrt{\frac{3\sinh^4(r)}{35\sinh^4(r)+8}}
\Bigg[\vert S_{0,r}\rangle-2\sqrt{\frac{2[\sinh^2(r)+1]}{\sinh^2(r)}}\vert S_{2,r}\rangle \nonumber \\
&\hspace{21ex} +2\frac{\sqrt{2}[\sinh^2(r)+1]}{\sqrt{3}\sinh^2(r)}\vert S_{4,r}\rangle\Bigg],
\end{align}
which is approximated by the superposition of squeezed number states;
\begin{align}
&\sqrt{\frac{3}{35}}
\Bigg[\vert S_{0,r}\rangle-2\sqrt{2}\vert S_{2,r}\rangle 
+2\sqrt{\frac{2}{3}}\vert S_{4,r}\rangle\Bigg].
\end{align}

\section{Proofs of Eq.~(22) and Eq.~(23)}\label{a6}
Here we prove Eq.~\eqref{eq1} and Eq.~\eqref{eq2} for completeness. Consider the Pascal matrix $P_ml$, defined as 
\begin{align}
P_{m,l}:=
\left\{
\begin{array}{ll}
{m \choose l} & \hbox{ when } m \ge l \\
0 & \hbox{ otherwise.}
\end{array}
\right.
\end{align}
 and its inverse, namely
\begin{align}
(P^{-1})_{m,l}=
\left\{
\begin{array}{ll}
(-1)^{m-l} {m \choose l} & \hbox{ when } m \ge l \\
0 & \hbox{ otherwise.}
\end{array}
\right.
\end{align}

The matrix element of the squeezing operator
$S(r)$ is given as
\begin{align}
S(r)_{j,m} &:= \frac{\sqrt{m! j !}}{(\cosh r)^{j+\frac{1}{2}}}
(\frac{\tanh r}{2})^{\frac{m-j}{2}}
\cos^2\frac{(j-m)\pi}{2}
F(r,m,j) \\
F(r,m,j) &:=
\sum_{k=\max(\frac{1}{2}(j-m),0)}^{\frac{j}{2}}
\frac{(-1)^k(\frac{1}{2}\sinh r)^{2k}}
{k! (j-2k)! [k+\frac{m-j}{2}]!}.
\end{align}
Using the above elements, we can express the squeezed photon number states as
\begin{align}
|S_{2m,r}\rangle &= \sum_{n=0}^{\infty} S(r)_{2n,2m} |2n\rangle \\
|S_{2m+1,e}\rangle &= \sum_{n=0}^{\infty} S(r)_{2n+1,2m+1} |2n\rangle.
\end{align}

Now, we consider the even number case.
Choosing $l=n-k$ and $j=2n$,
we have
\begin{align}
& |S_{2m,r}\rangle \nonumber \\
=&
\left(\frac{\sinh r}{\cosh r}\right)^{m}
\frac{\sqrt{(2m)!}}{2^{m}m!}
\sum_{l=0}^{m}
\frac{m!}{l! (m-l)!}
\frac{2^{2l}l!}{(2l)!}
\nonumber \\
& \cdot \bigg(\sum_{n=l}^{\infty}
\frac{\sqrt{(2n)!}}{{\cosh^{\frac{1}{2}} r}2^n n!}
\left(\frac{\sinh r}{\cosh r}\right)^{n}
\frac{n!(-1)^{n+l}}{(n-l)! \sinh^{2l} r} 
|2n\rangle \bigg).
\end{align}
We can now substitute this into Eq.~(\ref{alpha2m}). Since the coefficients in Eq.~(\ref{alpha2m}) are the inverse matrix of the Pascal matrix $P_{m,l}$, we have
\begin{align}
&|\alpha_{2l,r}\rangle \nonumber \\
=&
\frac{C_{2l,r} 2^{2l}l!}{(2l)!}
\sum_{n=l}^{\infty}
\frac{\sqrt{(2n)!}}{{\cosh^{\frac{1}{2}} r}2^n n!}
(\frac{\sinh r}{\cosh r})^{n}
\frac{n!(-1)^{n+l}}{(n-l)! \sinh^{2l} r} 
|2n\rangle.
\end{align}
Hence, we have
\begin{align}
\alpha_{2l,r|n} 
=&
\sqrt{\frac{\Gamma(\frac{1}{2})}{\Gamma(l+\frac{1}{2})}}
\Biggl( \frac{l!}{R^{l}} 
+
\frac{\Gamma(2l+\frac{1}{2})}{\Gamma(l+\frac{1}{2})}
\Biggr)^{-\frac{1}{2}}
\nonumber \\
&\cdot \frac{\sqrt{(2n)!}}{(1+R)^{\frac{1}{4}} 2^n n!}
(\frac{R}{1+R})^{\frac{n}{2}}
\frac{n!}{(n-l)! R^{l} } ,
\end{align}
where $R:= \sinh^2 r$.
We proceed to the analysis on the asymptotic behavior of the coefficients $\alpha_{2l,r|n}$.
When $n=xR$, 
using Stirling formula $n! \cong \sqrt{2\pi n}(\frac{n}{e})^n$, 
we have
$\frac{\sqrt{(2n)!}}{2^n n!}
\cong
(\pi x R)^{-1/4}$
and
$\frac{n!}{(n-l)! R^{l} } \to \frac{(xR)^l}{R^l}=x^l$.
Hence, we can conclude that $\alpha_{2l,r\vert n}\cong 
\frac{1}{R^{1/2}}
\sqrt{\frac{ \sqrt{\pi}}{\Gamma(2l+\frac{1}{2})}}
\pi^{-\frac{1}{4}}
e^{-\frac{x}{2}}
x^{l-\frac{1}{4}}
=\psi_{l-1/4}(x)/R^{1/2}$ in the asymptotic limit
because
$\Bigl( \frac{l!}{R^{l}} 
+
\frac{\Gamma(2l+\frac{1}{2})}{\Gamma(l+\frac{1}{2})}
\Bigr)^{-\frac{1}{2}}\to 
\sqrt{\frac{\Gamma(l+\frac{1}{2})}{\Gamma(2l+\frac{1}{2})}}$,
$(\frac{R}{1+R})^{\frac{n}{2}}
\to e^{-\frac{x}{2}}$,
and $(1+R)^{\frac{1}{4}}\cong R^{\frac{1}{4}}$.
So, we obtain \eqref{eq1}.

Next, we discuss the odd number case.
Choosing $l=n-k$ and $j=2n+1$,
we have
\begin{align}
&\vert\phi_{2m+1,R}\rangle
\nonumber \\
=&
(\frac{\sinh r}{\cosh r})^{m}
\frac{\sqrt{(2m+1)!}}{2^{m}m!}
\sum_{l=0}^{m}
\frac{m!}{l! (m-l)!}
\frac{2^{2l}l!}{(2l+1)!}
\nonumber \\
& \cdot \Bigg(
\sum_{n=l}^{\infty}
\frac{\sqrt{(2n+1)!}}{{\cosh^{\frac{3}{2}} r}2^n n!}
(\frac{\sinh r}{\cosh r})^{n}
\frac{n!(-1)^{n+l}}{(n-l)! \sinh^{2l} r} 
|2n+1\rangle \Bigg).
\end{align}
Now, we substitute the above relation in the place of $\vert\phi_{2m+1,R}\rangle$.
Since the coefficients in Eq.~(\ref{alpha2m}) are the inverse matrix of the Pascal matrix $P_{m,l}$,
we have
\begin{align}
&|\alpha_{2l+1,r}\rangle\nonumber \\
=&
\frac{C_{2l+1,r} 2^{2l}l!}{(2l+1)!}
\sum_{n=l}^{\infty}
\frac{\sqrt{(2n+1)!}}{{\cosh^{\frac{3}{2}} r}2^n n!}
\frac{\sinh^n r}{\cosh^n r}
\frac{n!(-1)^{n+l}}{(n-l)! \sinh^{2l} r} 
|2n+1\rangle.
\end{align}
Hence, we have
\begin{align}
&\alpha_{2l+1,r|n} \nonumber \\
=&
\sqrt{\frac{\Gamma(\frac{1}{2})}{\Gamma(l+\frac{1}{2})}}
\Biggl(
l! (2l+1 )R^{-l}
+
\frac{\Gamma (2l+\frac{1}{2})}{\Gamma (l+\frac{1}{2})}
(4l +1)
\Biggr)^{-\frac{1}{2}}\nonumber \\
&\cdot \frac{1}{(1+R)^{\frac{3}{4}}}
\frac{\sqrt{(2n+1)!}}{2^n n!}
(\frac{R}{1+R})^{\frac{n}{2}}
\frac{n!}{(n-l)! R^{l} } .
\end{align}
We proceed to the analysis on the asymptotic behavior of the coefficients $\alpha_{2l+1,r|n}$ in a similar fashion.
Using the approximation, we get
we have
\begin{align}
&
\alpha_{2l+1,r|n}\nonumber \\
\cong &
\sqrt{\frac{\Gamma(\frac{1}{2})}{\Gamma(2l+\frac{1}{2})}}
\cdot \frac{\sqrt{2n+1}\sqrt{(2n)!}}{(4l +1)^{\frac{1}{2}}(1+R)^{\frac{3}{4}}2^n n!}
(\frac{R}{1+R})^{\frac{n}{2}}
\frac{n!}{(n-l)! R^{l} }
\nonumber \\
=& \frac{1}{
R^{1/2}}\psi_{l+\frac{1}{4}}(x).
\end{align}
Thus, we obtain 
$\alpha_{2l+1,r|n} \cong 
\frac{1}{R^{1/2}}\psi_{l+\frac{1}{4}}(x)$
as $r \to \infty$.
So, we obtain \eqref{eq1}.

\section{Experimental Scheme for Production of Suboptimal State}\label{a7}
We include some details of the derivation of the scheme to generate superpositions of squeezed state. We begin with the two-mode squeezed vacuum (TMSV) state, namely $\vert TMSV\rangle=\sqrt{1-q^2}\sum_{n=0}^{\infty}\vert n\rangle_s\otimes\vert n\rangle_h$, with $q=\tanh(r_2)$ relating the squeezed vacuum to the two mode squeezing parameter $r_2$. Since we are performing a four photon coincidence measurement, we write the post-measurement state as 
\begin{align}
\vert\phi\rangle\propto\langle0\vert_h(\frac{a}{2}+\beta_1)(\frac{a}{2}+\beta_2)(\frac{a}{2}+\beta_3)(\frac{a}{2}+\beta_4)\vert\text{TMSV}_q\rangle.
\end{align}
This arises from displacing the annihilation operator $D(\alpha)aD^{\dagger}(\alpha)=a+\alpha$ corresponding to each measured mode. This measurement produces a heralded state at the idler port of the TMSV that is given by $\vert\psi\rangle=N^{-1}\sum_{k=0}^{4}\phi_k\vert k\rangle$, where $\phi_0:=\beta_1\beta_2\beta_3\beta_4$, $\phi_1:=\frac{q}{2}(\beta_1\beta_2\beta_3+\beta_1\beta_2\beta_4+\beta_1\beta_3\beta_4+\beta_2\beta_3\beta_4)$, $\phi_2:=\frac{q^2\sqrt{2}}{4}(\beta_1\beta_2+\beta_1\beta_3+\beta_1\beta_4+\beta_2\beta_3+\beta_2\beta_4+\beta_3\beta_4)$, $\phi_3:=\frac{q^3\sqrt{3!}}{8}(\beta_1+\beta_2+\beta_3+\beta_4)$ and $\phi_4:=\frac{q^4\sqrt{4!}}{2^4}$. As stated in the main text, making the choice $\beta_1\beta_2=\beta_3\beta_4$ and $\beta_1+\beta_2=-(\beta_3+\beta_4)$ eliminates the odd sectors $\phi_1=0$ and $\phi_3=0$. Furthermore, the choice $r_2=2$, $\beta_1=0.343824$ and $\beta_2=2.59058$ produces a state with $\approx 99.94\%$ fidelity with the desired subopt
 imal sta
 te with arbitrary squeezing $r$.

%


\end{document}